# Joint brain tumor segmentation from multi MR sequences through a deep convolutional neural network

Farzaneh Dehghani[1], Alireza Karimian[1], Hossein Arabi[2]

[1]Department of Biomedical Engineering, Faculty of Engineering, University of Isfahan, Isfahan, Iran

[2]Division of Nuclear Medicine and Molecular Imaging, Department of Medical Imaging, Geneva University Hospital, CH-1211 Geneva 4, Switzerland.

**Abstract**

Purpose

Brain tumor segmentation is highly contributive in diagnosing and treatment planning. The manual brain tumor delineation is a time-consuming and tedious task and varies depending on the radiologist's skill. Automated brain tumor segmentation is of high importance, and does not depend on either inter or intra-observation. The objective of this study is to automate the delineation of brain tumors from the FLAIR, T1-weighted, T2-weighted, and T1-weighted contrast-enhanced MR sequences through a deep learning approach, with a focus on determining which MR sequence alone or which combination thereof would lead to the highest accuracy therein.

Method

The BRATS-2020 challenge dataset, containing 370 subjects with four MR sequences and manually delineated tumor masks, are applied to train a residual neural network. This network is trained and assessed separately for each one of the MR sequences (single-channel input) and any combination thereof (dual- or multi-channel input).

Results

The quantitative assessment of the single-channel models reveals that the FLAIR sequence would yield higher segmentation accuracy compared to its counterparts with a $0.77 \pm 0.10$ Dice index. As to considering the dual-channel models, the model with FLAIR and T2W inputs yield a $0.80 \pm 0.10$ Dice index, exhibiting higher performance. The joint tumor segmentation on the entire four MR sequences yields the highest overall segmentation accuracy with a $0.82 \pm 0.09$ Dice index.

Conclusions

The FLAIR MR sequence is considered the best choice for tumor segmentation on a single MR sequence, while the joint segmentation on the entire four MR sequences would yield higher tumor delineation accuracy.

## 1- Introduction

Brain tumor is defined as the abnormal growth of cells or the central spinal canal [1]. The most common brain tumors are gliomas, typically categorized into High-Grade Gliomas (HGG) and Low-Grade Gliomas (LGG) [2]. Clinicians obtain the necessary information regarding tumor progression, evolution, and response to the therapy by acquiring data from different medical imaging modalities. Magnetic Resonance Imaging (MRI) is an effective and sensitive imaging modality for the task of the tumor, lesion, tissue, and disease identification/characterization owing to its high soft-tissue contrast compared to other modalities such as CT imaging [3-9].

The brain tumor segmentation process consists of delineation discrimination of the brain tumor tissues from surrounding normal tissues. Accurate and reliable brain tumor segmentation is highly contributive in disease diagnosis, monitoring, and treatment planning [10]. Although manual brain tumor segmentation is frequently performed by radiologists, it is a highly cumbersome and time-consuming task. This type of brain segmentation from MR images is prone to intra-and inter-rater variations. Consequently, an accurate and reliable automated brain tumor segmentation method is highly required in clinical settings [11, 12].

Automated tissue/organ delineation/identification from anatomical MR images could be conducted through different techniques, consisting of atlas-based, shape-based averaging, principle component analysis (PCA), and active contour approaches [13, 14]. The atlas-based method requires a number of templates/atlases with ground-truth tissue/organ label maps to be aligned/deformed to the target MR images [15, 16]. The shape-based averaging method relies on the atlas images and the signed distant map calculation for each label in the atlas dataset to identify the target tissue/organ on the target MR [17]. The PCA-based methods tend to reduce the dimensionality of the input data or variability of the target tissue/organ, while, retaining the significant or representative variations to identify the target tissue/organ [18]. The active contour method is based on the energy minimization of a deformable spline, which is influenced by the image forces and some predefined constraints to iteratively define the target tissue/organ contour [19]. These approaches have exhibited high degrees of success in deep learning vs. conventional methods in this context [20-22].

In recent years, some researchers, by focusing on brain tumor segmentation, issue have developed innovative deep learning-based approaches with highly improved automatic brain tumor identification and delineation. Reasearchers in [23] proposed a complex convolutional neural network (CNN) architecture for the task of brain tumor segmentation from MR images (MICCAI brain tumor segmentation challenge (BRATS 2013)), wherein a cascade of two sequential CNN modules were employed for a robust tumor delineation. Researchers in [11] run a similar study on brain tumor segmentation by applying a CNN model together with intensity normalization and data augmentation to enhance the accuracy of the tumor segmentation compared to that of previous studies. They obtained a Dice index of 0.88, 0.83, and 0.77 for the entire, core, and enhancing tumor regions, respectively. Rundo et al. [6] proposed a semi-automatic brain tumor segmentation method. In this study, they applied an unsupervised Fuzzy C-Means clustering technique to segment the target and calculate the lesion volume automatically.

A model is proposed in [24] based on U-Net architecture for brain tumor detection and segmentation on 220 MR images from the BRATS 2015 dataset and reported a Dice index of 0.86. In a similar sense, a U-Net architecture for automated brain tumor segmentation, and radiomics-based survival prediction is proposed in [25], where 0.896, 0.797, and 0.732 Dice scores are obtained for the whole tumor, tumor core, and enhancing tumor regions, respectively on the BRATS 2015 dataset. A 3D U-Net model is proposed for automated brain tumor segmentation and patient survival prediction by applying the BRATS 2018 dataset in [26], where the gross tumor volume obtained is then fed into a CNN network to classify the entire tumor volume into the tumor core, enhancing tumor, and peritumoral edema. In a similar study, researchers in [27] proposed a two-stage framework for automatic brain tumor segmentation, where, the whole brain tumor is delineated in the first stage through the random forest classifier and dense conditional random field, and the voxels within the obtained mask are classified into different tumor/tissue types in the second stage,.

Different MR sequences could be applied for imaging brain/tumor in a single acquisition session. The segmentation of tumors could be applied separately or jointly on different MR sequences. The BRATS dataset applied in this study consists of brain MR images in the four, FLAIR, T1W, T2W, and T1W contrast-enhanced (T1Wc) sequences.

The objective of this study is to develop a deep learning model for automated segmentation of the brain tumor from the different MR sequences. To fulfill this objective, enough deep learning models will be developed to perform brain tumor delineation on each MR sequence (single-channel input) and any combination therein (multi-channel input) to determine the most efficient MR sequences in an independent manner for this purpose. Introducing the most effective models (single-, dual-, and multiple-channel inputs) and their corresponding MR sequences for the task of automated brain tumor delineation is the primary focus of this study.

## 2- Methods and materials

The automatic brain tumor segmentation from the different MR sequences is assessed for different scenarios including training of deep learning models with single-, dual-, triple-, and quad-channel input/inputs. Different combinations of the MR sequences as the input images to the deep learning models are assessed to identify the most effective models in terms of accuracy in brain tumor segmentation.

### 2-1- Dataset

The publicly available BRATS 2020 dataset containing 370 subjects with HGG and LGG brain tumors, each including T1-weighted (T1W), T1-weighted imaging with gadolinium-enhancing contrast (T1Wc), T2-weighted (T2W), and FLAIR images together with a single manually defined tumor mask is applied in this study. The entire MR sequences being segmented together into necrotic/non-enhancing tumor, peritumoral edema, and enhancing tumor tissues by experienced radiologists (there is only one tumor mask for each subject/patient or the four MR sequences).

### 2-2- Pre-processing

Prior to the training of the deep learning models, the MR sequences are cropped as to only contain the tumor and the background tissues. The MR sequences are equally cropped to remove the background-air and the irrelevant tissues. Among all the segmented images of the patients in the BRATS dataset, the largest tumor size is identified in three x, y, and z directions. Because the input image size fed into the deep learning models should be a multiplication of 8, first, the entire MR images (T1W, T1Wc, T2W, and FLAIR sequences) were cropped to a 144×128×96 mm$^3$ sub-volume which encompassed the whole tumor and the neighboring background tissues. Then, the image intensities of the entire MR images (cropped sub-volumes) were normalized within a 0-1 range through division by the maximum values of each MR sequence, separately.

**2-3- Convolutional Neural Network**

The ResNet architecture, implemented in the NiftyNet platform, was adopted to build the different tumor delineation models. The NiftyNet is a TensorFlow-based machine learning platform created for medical image analysis/processing purposes [28].

The 310 MRI images in four T1W, T1Wc, T2W, and FLAIR sequences are applied as the inputs for training the CNN Model. The input image size is 144×128×96 and the architecture is of HighRes3DNet. The ResNet is a CNN architecture, consisting of 20 layers. In the first seven layers, the low-level features like edges are extracted from the input data through a 3×3×3-voxel convolutional kernel; in the next seven layers, a dilated convolutional kernel by a factor of two is applied to encode the medium-level features from the input and the last six layers apply a dilated convolutional kernel with a factor of four, to extract the high-level features. A batch normalization and elementwise rectified linear unit (ReLU) is connected to the convolutional layers in the residual blocks. A residual block is constructed by connecting every two convolutional layers using a residual connection [29, 30].

Each one of the 13 CNN models is trained separately. The inputs of these models consist of the 4 T1W, T1Wc, T2W, and FLAIR single-channel sequences, 6 T1W + T1Wc, T1W + T2W, T1W + FLAIR, T1Wc + T2W, T1Wc + FLAIR, and T2W + FLAIR dual-channel sequences, and 3 T1W + T1Wc + FLAIR, T1W + T2W + FLAIR, and T1W + T1Wc + T2W + FLAIR multi-channel sequences. Here, the Adaptive Moment Estimation (Adam) and Dice-NS are applied as the optimizer and loss functions, respectively. The learning rate is 0.01 at 15 batch size. The number of training iterations was 10000, 12000, and 20000 for each of the single-, dual-, and multi-channel models, respectively. To improve the performance of this proposed brain tumor segmentation, the dual-channel and multi-channel CNN is applied to the sub-volume and the results are compared with that of the single-channel. In this experiment, each MRI image is trained and optimized separately, enabling the network to determine the image with the best performance for brain tumor segmentation.

**2-4- Validation**

A total of 13 different ResNet models were trained and 60 subjects (each involving four MR sequences) as an external test dataset were employed for the evaluation of the models. To evaluate the performance of these models, the accuracy of the resulting tumor masks was assessed through standard segmentation metrics. The accuracy of segmentation is determined by applying the sensitivity, precision, Dice Similarity coefficient, Jaccard index, and Hausdorff Distance parameters, calculated through Eqs. (1-5) [31]:

$$Sensitivity(S, M) = \frac{|S \cap M|}{|M|} \quad (1)$$

$$Precision(S, M) = \frac{|S \cap M|}{|S|} \quad (2)$$

$$Dice\ similarity\ coefficient = \frac{2 \times |S \cap M|}{|S| + |M|} \quad (3)$$

$$Jaccard\ index = \frac{|S \cap M|}{|S \cup M|} \quad (4)$$

$$Hausdorff\ Distance(S, M) = \max(h(A, M), h(M, S)) \quad (5)$$

$$, h(S, M) = \max_{s \in S} \min_{m \in M} \|s - m\|$$

where, S and M are defined as the automatically and the manually segmented region, respectively. Sensitivity is defined as the ratio of correctly labeled tumor area to the entire area of the reference tumor. Precision is defined as the proportion of the area/volume that the model correctly identified the tumor to the total area of the tumor delineated by the model. The Dice similarity coefficient and the Jaccard index are applied in comparing the degree of similarity or difference between the reference segmented tumor and the tumor segmented through this model. The Hausdorff Distance is applied to measure the distance between the reference segmented tumor of the segmented tumor through the deep learning models.

**3- Results**

The performance of the single-, dual-, and multi-channel deep learning models is assessed through different standard segmentation metrics. Representative binary masks of the brain tumor segmented by the single-channel and multi-channel models are illustrated in Fig. 1.

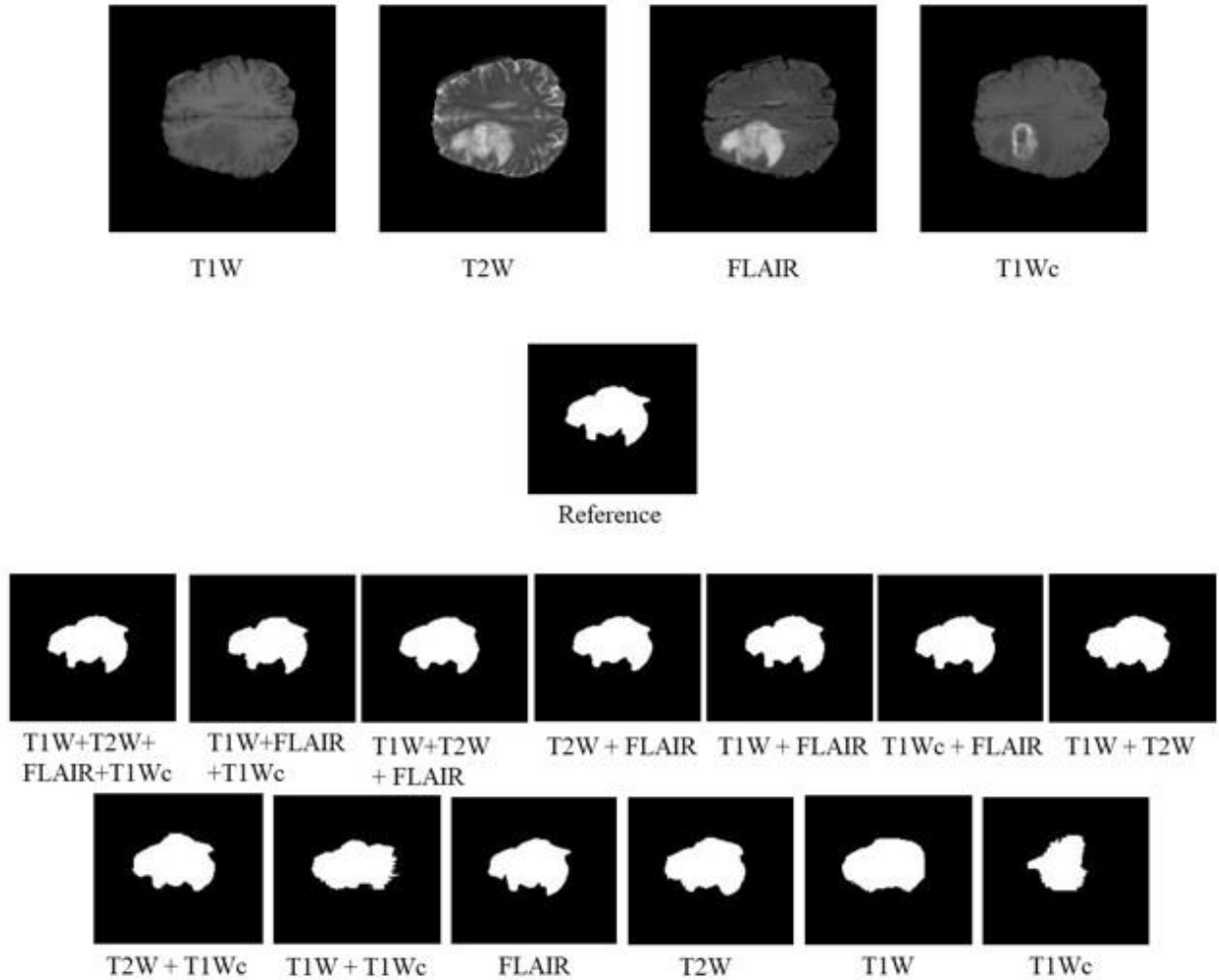

**Fig. 1.** Representative binary mask of the brain tumor segmented by the single-channel and multi-channel models.

The details of the sensitivity and precision results for the different deep learning models are tabulated in Table 1, and the results of the Dice Similarity Coefficient, Jaccard index, and Hausdorff Distance are tabulated in Table 2.

**Table 1.** The results of sensitivity and precision for the different single- and multi-channel deep learning models for brain tumor segmentation.

| Mode | Sensitivity (mean ± SD) [min, max] | Precision (mean ± SD) [min, max] |
|---|---|---|
| **FLAIR+T1W+T2W+T1Wc** | 0.80 ± 0.13 [0.48, 0.98] | 0.87 ± 0.09 [0.48, 0.98] |
| **FLAIR+T1W+T2W** | 0.76 ± 0.17 [0.31, 0.98] | 0.90 ± 0.08 [0.46, 0.98] |
| **FLAIR+T1W+T1Wc** | 0.77 ± 0.15 [0.41, 0.97] | 0.84 ± 0.10 [0.34, 0.98] |
| **T2W + FLAIR** | 0.82 ± 0.13 [0.44, 0.98] | 0.80 ± 0.12 [0.27, 0.97] |
| T1W + FLAIR | 0.81 ± 0.13 [0.45, 0.98] | 0.78 ± 0.12 [0.26, 0.95] |
| T1Wc + FLAIR | 0.85 ± 0.12 [0.57, 0.99] | 0.73 ± 0.15 [0.22, 0.94] |
| T1W+T2W | 0.73 ± 0.18 [0.27, 0.98] | 0.82 ± 0.13 [0.36, 0.97] |
| T2W + T1Wc | 0.79 ± 0.14 [0.43, 0.97] | 0.76 ± 0.14 [0.22, 0.94] |
| T1W + T1Wc | 0.72 ± 0.13 [0.34, 0.92] | 0.61 ± 0.15 [0.20, 0.83] |
| FLAIR | 0.83 ± 0.12 [0.55, 0.99] | 0.74 ± 0.14 [0.24, 0.92] |
| T1W | 0.79 ± 0.16 [0.19, 0.97] | 0.69 ± 0.12 [0.29, 0.89] |
| T2W | 0.73 ± 0.18 [0.31, 0.98] | 0.77 ± 0.17 [0.27, 0.97] |
| T1Wc | 0.68 ± 0.121 [0.09, 0.94] | 0.60 ± 0.16 [0.07, 0.83] |

**Table 2.** The results of Dice, Jaccard, and Hausdorff Distance for the different single- and multi-channel deep learning models for brain tumor segmentation.

| Mode | Dice (mean ±SD) [min, max] | Jaccard (mean ± SD) [min, max] | Hausdorff Distance (mean ±SD) [min, max] |
|---|---|---|---|
| FLAIR+T1W+T2W+T1Wc | 0.82 ± 0.09 [0.48, 0.94] | 0.71 ± 0.12 [0.32, 0.89] | 3.11 ± 0.51 [1.96, 4.87] |
| FLAIR+T1W+T2W | 0.81 ± 0.12 [0.45, 0.95] | 0.70 ± 0.15 [0.29, 0.90] | 3.01 ± 0.56 [1.80, 4.81] |
| FLAIR+T1W+T1Wc | 0.79 ± 0.10 [0.41, 0.94] | 0.67 ± 0.13 [0.26, 0.89] | 3.31 ± 0.63 [2.28, 5.22] |
| T2W + FLAIR | 0.80 ± 0.10 [0.38, 0.93] | 0.68 ± 0.12 [0.24, 0.87] | 3.28 ± 0.53 [2.30, 5.01] |
| T1W + FLAIR | 0.79 ± 0.10 [0.37, 0.94] | 0.66 ± 0.12 [0.23, 0.90] | 3.50 ± 0.74 [2.24, 6.63] |
| T1Wc + FLAIR | 0.77 ± 0.11 [0.33, 0.94] | 0.63 ± 0.13 [0.20, 0.88] | 3.68 ± 0.67 [2.37, 5.90] |
| T1W + T2W | 0.76 ± 0.13 [0.30, 0.93] | 0.63 ± 0.16 [0.18, 0.87] | 3.25 ± 0.55 [2.37, 5.19] |
| T2W + T1Wc | 0.76 ± 0.11 [0.29, 0.90] | 0.63 ± 0.13 [0.17, 0.82] | 3.64 ± 0.51 [2.84, 5.25] |
| T1W + T1Wc | 0.66 ± 0.13 [0.25, 0.84] | 0.50 ± 0.13 [0.14, 0.72] | 4.01 ± 0.51 [3.13, 5.35] |
| FLAIR | 0.77 ± 0.10 [0.34, 0.94] | 0.64 ± 0.12 [0.20, 0.89] | 3.65 ± 0.64 [2.30, 5.18] |
| T1W | 0.73 ± 0.13 [0.23, 0.88] | 0.59 ± 0.14 [0.13, 0.78] | 3.60 ± 0.50 [2.84, 5.18] |
| T2W | 0.73 ± 0.15 [0.32, 0.91] | 0.60 ± 0.16 [0.19, 0.83] | 3.49 ± 0.51 [2.52, 5.26] |
| T1Wc | 0.62 ± 0.17 [0.08, 0.84] | 0.46 ± 0.15 [0.04, 0.78] | 3.98 ± 0.56 [2.51, 5.48] |

The results of this metric on all patient's MRI images are illustrated in Figs. (2, 3, and 4) to facilitate the assessment of the Dice performance in different modalities more accurately.

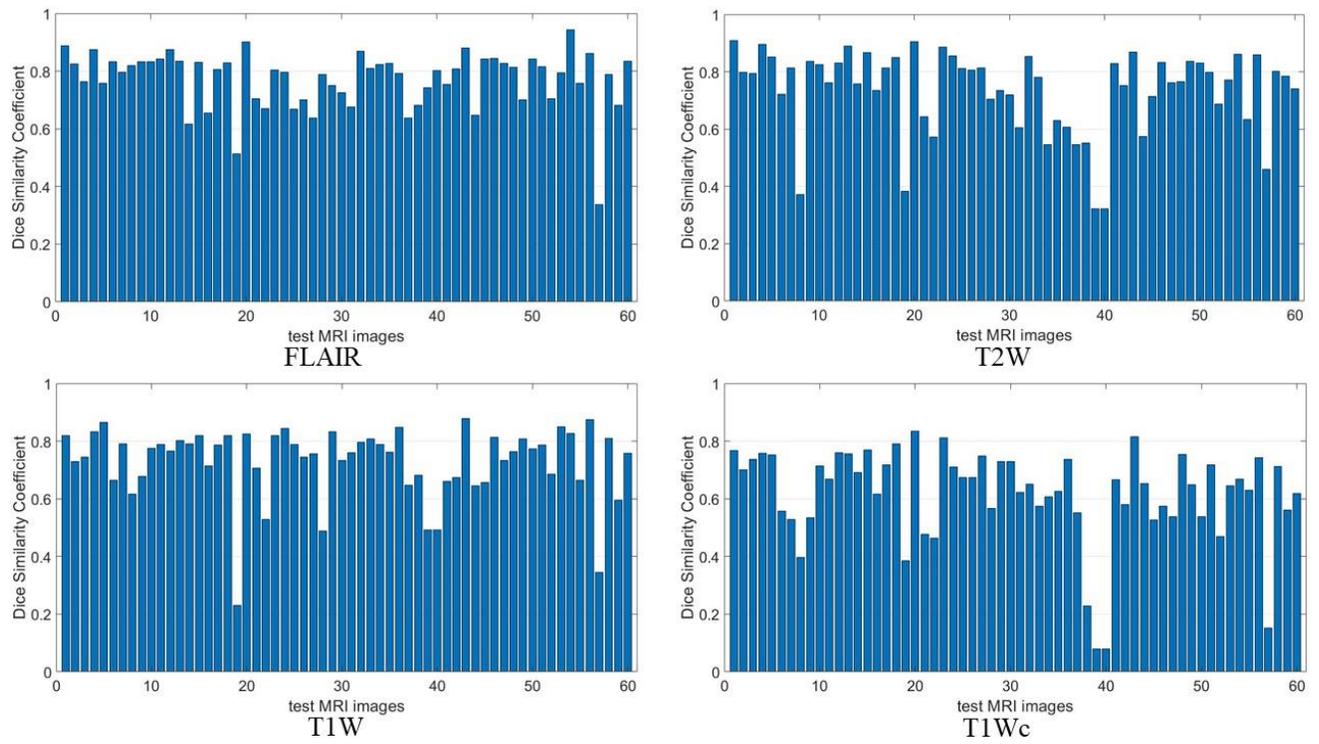

**Fig. 2.** The results of Dice for single-channel MRI images in the test dataset

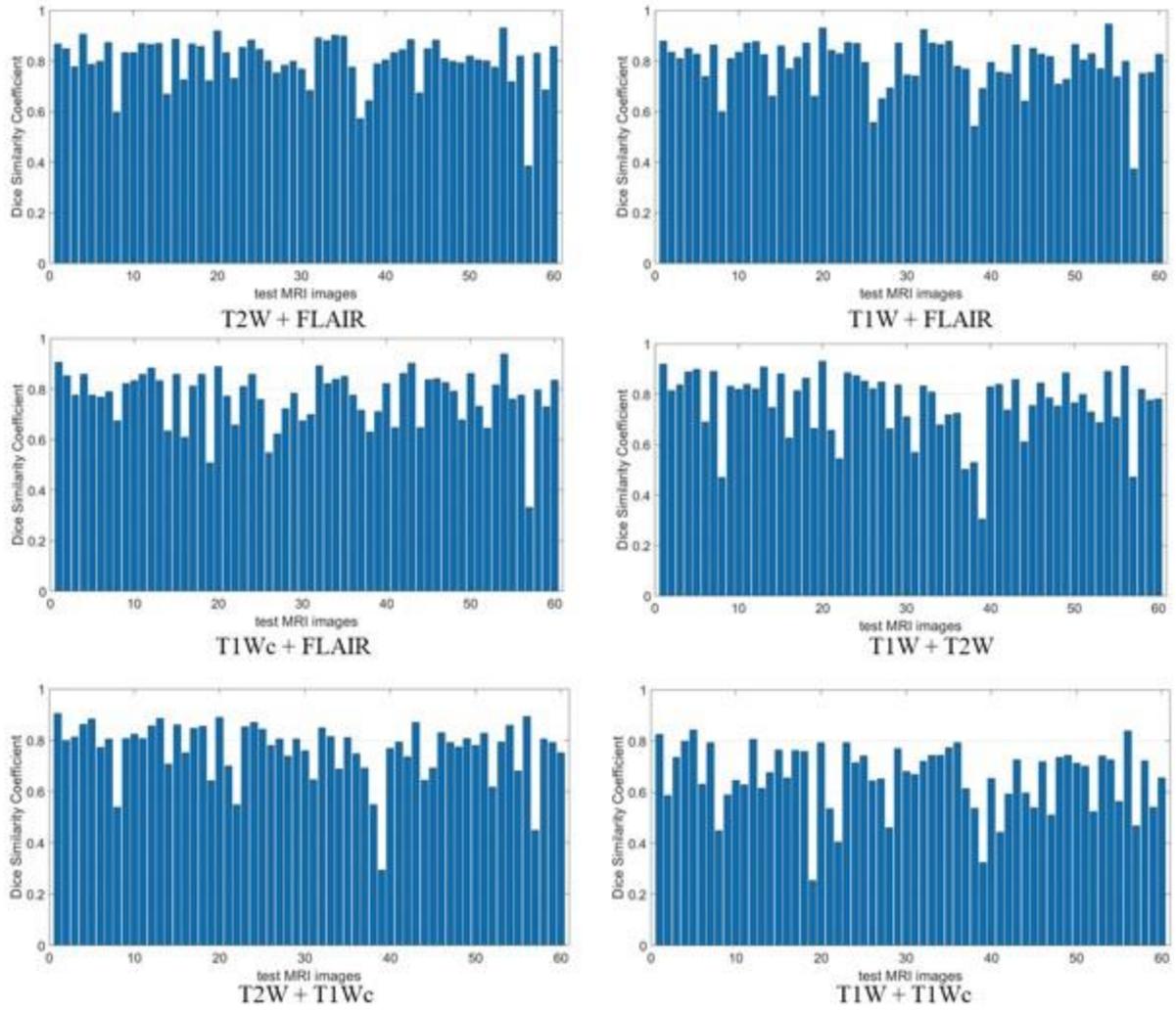

**Fig. 3.** The results of Dice for dual-channel MRI images in the test dataset

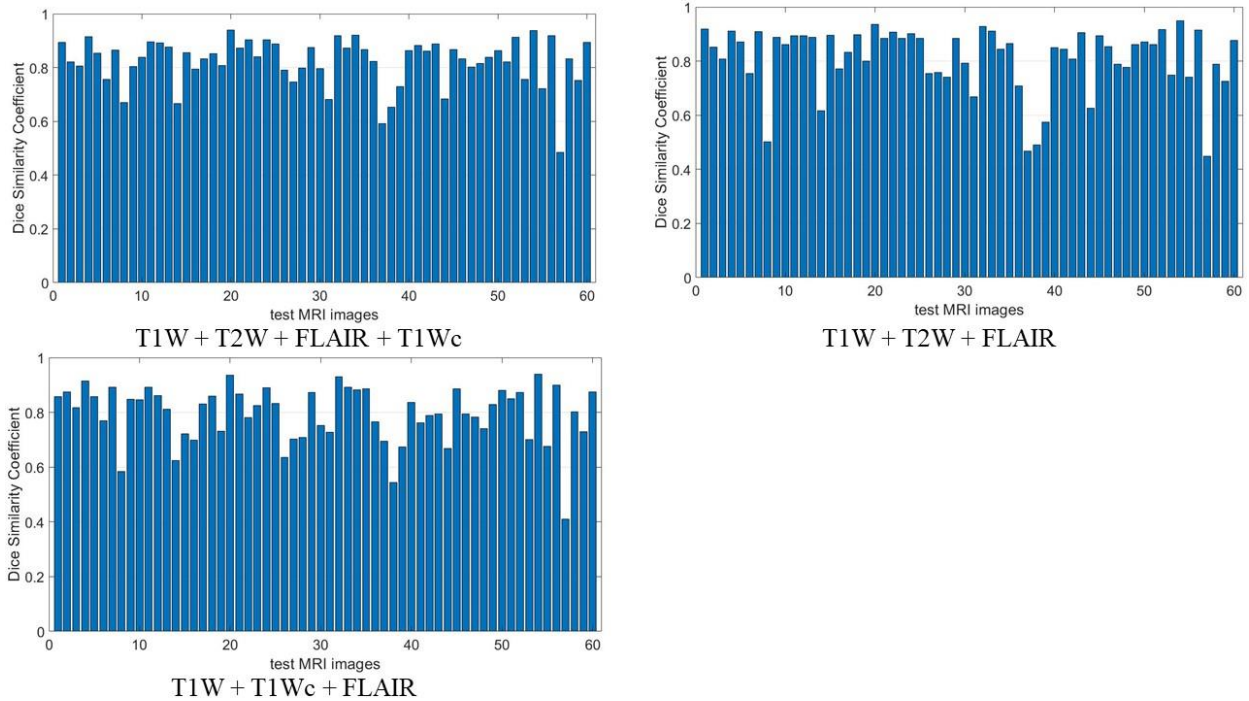

**Fig. 4.** The results of Dice for multi-channel MRI images in the test dataset

### 4- Discussion

The objective of this study is to propose deep learning models for reliable brain tumor segmentation. To this end, the performance of the different single- and multi-channel deep learning models was evaluated for the task of the automated brain tumor segmentation from the different MR sequences.

The results of sensitivity and precision reveal that this proposed CNN approach is highly reliable for brain tumor segmentation, indicating the acceptable performance of T1Wc + FLAIR and T1W + T2W + T1Wc + FALIR, respectively, Table 1. As observed in this table, the T1W, T2W, and T1Wc sequences' performance is improved when fused with the FLAIR sequence.

To provide more accurate brain tumor segmentation, Dice Similarity Coefficient, Jaccard index, and Hausdorff distance are calculated. The results obtained from Dice reveal that the performance of all dual/multi-channel CNNs is more reliable than that of the single-channel CNNs, except for T1W + T1Wc and FLAIR. The results obtained from Dice and Jaccard indicate that T1W + T2W + T1Wc + FLAIR and T1W + T2W + FLAIR CNNs are more reliable than that of the other sequences for brain tumor segmentation. Considering the single- and dual-channel models, the FLAIR and T2W+FLAIR sequences as input images resulted in relatively superior outcomes compared to the other single- and dual-channel models.

The results of the Dice Similarity Coefficient on all patient's MRI images indicate that multi-channel CNNs are more promising approaches for automated brain tumor segmentation. The main reason for the improved results in joint segmentation is that the different MR sequences provide different representations of the same brain tumor. Thus, the multi-channel models would benefit from the complementary information that exists in the different MR sequences, and as a result, would lead to a more accurate brain tumor delineation. As observed in Fig. 2, the Dice indices when using T1Wc sequence as a single input were inferior to those of the dual-channel model with T1Wc + FLAIR as input (Fig. 3).

As to the results obtained from the single-channel models, it is deduced that the FLAIR sequence bears the most effective information related to the discrimination of the tumor tissue from the back-ground healthy tissues. The T1W, T2W, and T1Wc sequences contain less discriminative features for identification of the brain tumors, while each one of these MR sequences provides a unique set of image features to distinguish the tumor from the normal tissues. These image features are not common in the four MR sequences, therefore, the multi-channel model took advantage of all four MR images resulted in the highest tumor segmentation accuracy. Because tumor representation varies in different MR sequences, a combination of these MR sequences would provide a synergy for optimal brain tumor identification.

More assessments conducted through the BRATS dataset revealed that the tumor region is not clear in some MRI images, as a result, the brain tumor segmentation error is inevitable in these MR images (Fig. 5).

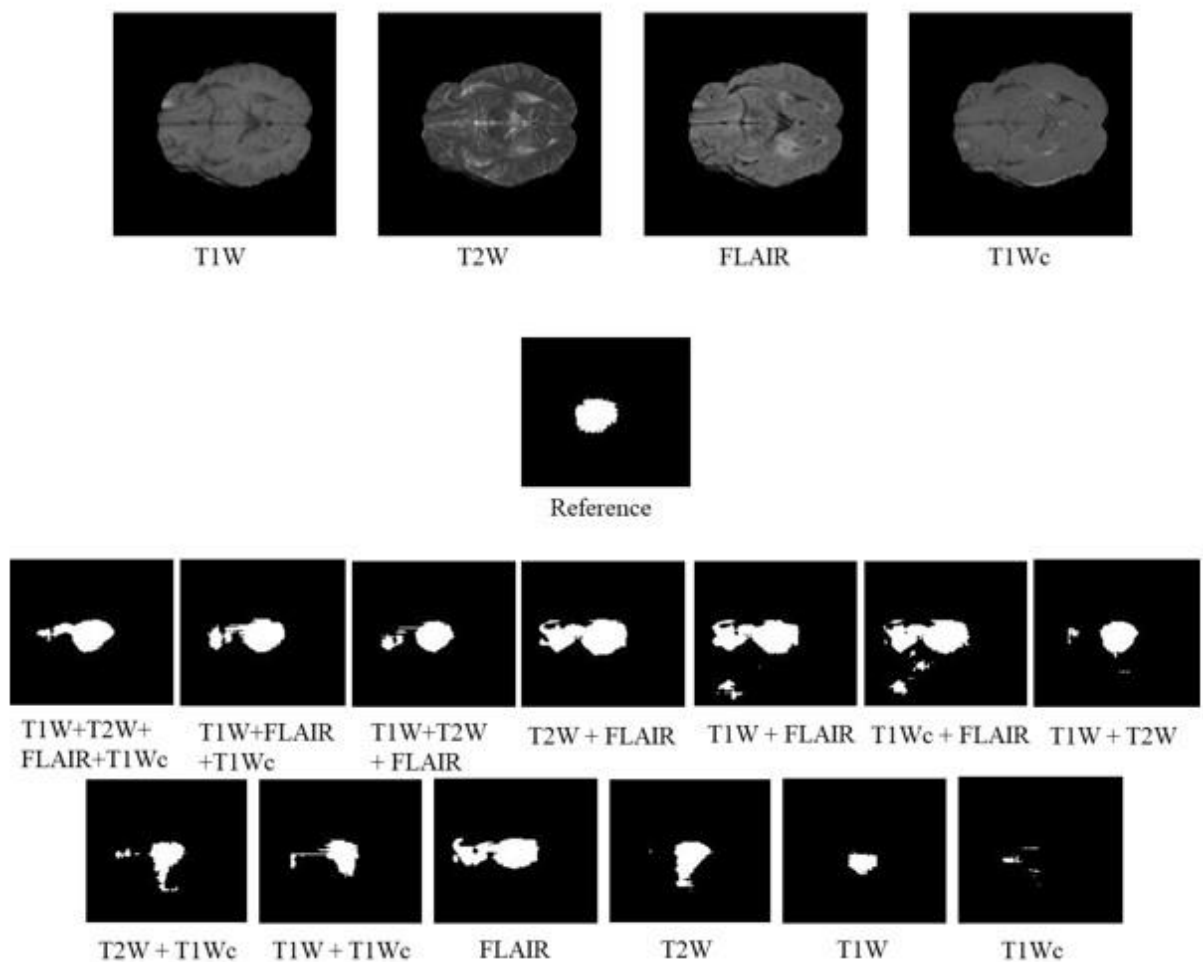

**Fig. 5.** An example of MR image where the deep learning models for brain tumor segmentation failed. The results obtained from Dice are 0.34, 0.15, 0.45, and 0.29 for T1W, T1Wc, T2W, and FLAIR, respectively. For the joint segmentation, the results of Dice are 0.47, 0.45, 0.38, 0.37, and 0.33 for T1W + T2W and T1W + T1Wc, T2W + T1Wc, T2W + FLAIR, T1W + FLAIR, and T1Wc + FLAIR, respectively. For T1W + T2W + FLAIR + T1Wc, T1W + T2W + FLAIR, and T1W + T1Wc + FLAIR the results of Dice are 0.48, 0.45, and 0.41 respectively. Although the results are imperfect, dual, and multi-channel sequences outperform the single-channel sequences.

For more comprehensive assessments, the performance of multi MR sequences (3 or 4 inputs) is assessed for brain tumor segmentation. Although the results of multi MR sequences are enhanced, the extended imaging time is still a real challenge. The objective of this study would be accomplished through the best dual-channel model with T2W + FLAIR sequences as input, where the results of the brain tumor segmentation is increased up to 0.80 in terms of Dice index.

## 5- Conclusion

MRI imaging is a time-consuming and cumbersome procedure to which the patients are subject to. Identifying MRI sequences with remarkable performance is of great concern for clinical applications. The objective of this study is to find MRI sequences with the best performance for brain tumor segmentation as to avoid unnecessary assessment of imperfect sequences for this purpose.

The results reveal that the joint tumor segmentation on multi- and dual-channel sequences outperforms the segmentation on single-channel MR sequences. The obtained results indicate that T1W + T2W + FLAIR + T1Wc, T1W + T2W + FLAIR, and T2W + FLAIR are significantly reliable for automatic brain tumor segmentation. The results obtained from the segmentation of single-channel sequences indicate that the FLAIR sequence is more reliable for brain tumor segmentation than other sequences.

**Acknowledgments**

The authors would like to thank the University of Isfahan and the Avicenna Center of Excellence (ACE) for their support.

**Funding**

This research is financed by the University of Isfahan, subject to grant number 9912011.

**Conflicts of interest**

There is no conflict of interest in this work.